\documentclass[prl,twocolumn,showpacs]{revtex4}
\usepackage{amsmath}
\usepackage[dvips,xdvi]{graphicx}
\usepackage{amssymb}
\usepackage{epsfig}

\catcode`\Š = \active \catcode`\š = \active \catcode`\Ÿ = \active
\catcode`\€ = \active \catcode`\… = \active \catcode`\† = \active
\catcode`\§ = \active \catcode`\Ž = \active \catcode`\ = \active
\catcode`\' = \active \catcode`\™ = \active \catcode`\ = \active
\catcode`\¿ = \active \catcode`\˜ = \active \catcode`\' = \active
\defŠ{\"a} \defš{\"o} \defŸ{\"u} \def€{\"A} \def…{\"O} \def†{\"U} \def§{\ss} \defŽ{\'{e}}
\def{\`{e}} \def'{\"{e}} \def™{\^{o}} \def{\^{e}} \def¿{\o} \def˜{\`{o}} \def'{\'{i}}

\begin{document}
\title{Coherent Molecular Optics using Ultra-Cold Sodium Dimers}
\author{J.R. Abo-Shaeer\footnote{Contact Info: jamil@mit.edu}, D.E. Miller, J.K. Chin, K. Xu, T. Mukaiyama, and W. Ketterle\footnote{Website: cua.mit.edu/ketterle\_group}}

\affiliation{Department of Physics, MIT-Harvard Center for
Ultracold Atoms, and Research Laboratory of Electronics, MIT,
Cambridge, MA 02139}
\date{\today}

\begin{abstract}
Coherent molecular optics is performed using two-photon Bragg
scattering. Molecules were produced by sweeping an atomic
Bose-Einstein condensate through a Feshbach resonance. The
spectral width of the molecular Bragg resonance corresponded to an
instantaneous temperature of 20 nK, indicating that atomic
coherence was transferred directly to the molecules. An
autocorrelating interference technique was used to observe the
quadratic spatial dependence of the phase of an expanding
molecular cloud. Finally, atoms initially prepared in two momentum
states were observed to cross-pair with one another, forming
molecules in a third momentum state.  This process is analogous to
sum-frequency generation in optics.

\end{abstract}


\maketitle

Similar to the field of optics, where the high intensity and
coherence of lasers allowed for the observation of effects such as
frequency doubling and wave-mixing, atom optics has benefited
greatly from the realization of Bose-Einstein condensates (BEC).
High phase-space density (atoms per mode) and a uniform phase
\cite{sten99brag,hagl99coh} give the condensate its laser-like
qualities.  Although not fundamentally required
\cite{moor00atom,kett01fermi}, BEC has led to the observation of
such phenomena as four-wave mixing \cite{deng99}, matter wave
amplification \cite{inou99mwa,kozu99amp}, and atom number
squeezing \cite{orze01}.

The current state of molecular optics is similar to atom optics
prior to the realization of BEC.  Diffraction and interferometry
of thermal molecular beams has been demonstrated
\cite{bord94,chap95,hack03,cork98}, yet monoenergetic beams lack
the density necessary to observe nonlinear effects. However,
recent experiments using Feshbach resonances have demonstrated the
conversion of degenerate atomic bosons \cite{durr03mol,
herb03cs_mol, xu03na_mol} and fermions \cite{rega03mol,joch03BEC,
zwie03molBEC, cubi03, stre03} into ultracold molecules.  These
sources have the potential to greatly advance molecular optics.
Furthermore, atom-molecule coupling can be studied as the first
steps towards ``superchemistry", where chemical reactions are
stimulated via macroscopic occupation of a quantum state
\cite{hein00super}.
\begin{figure}[h]
\includegraphics[width=75mm]{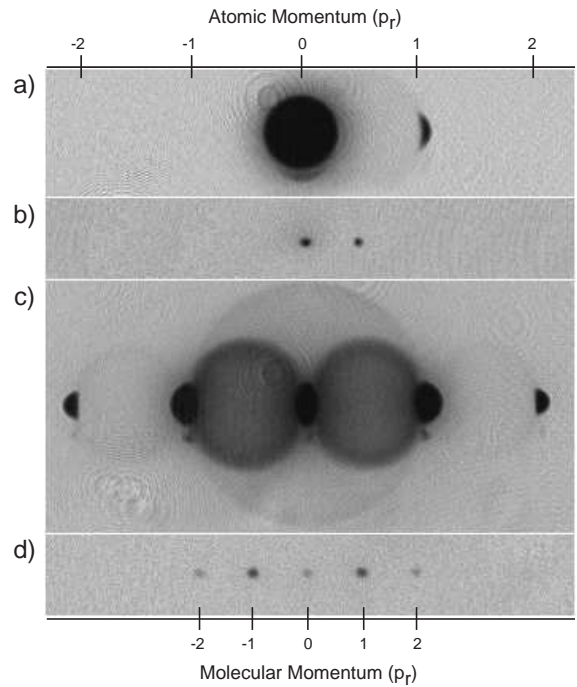} \caption{Bragg
diffraction of (a) atoms and (b) molecules.  Scattered particles
recoil with identical momenta, $p_{r}$. However, during ballistic
expansion diffracted molecules expand with half the velocity of
atoms, due to their doubled mass. The pulse duration in each image
was 200 $\mu s$.  A shorter pulse (5 $\mu s$) populated multiple
(c) atomic and (d) molecular momentum states. The halos in (c) are
due to collisions between different momentum states \cite{chik00}.
The time-of-flight in each image is 17 ms.} \label{fig:Figure1}
\end{figure}

The coherent manipulation of atomic matter waves via stimulated
light scattering has been crucial in the development of nonlinear
atom optics (for a review see \cite{rols02_nature}).  Here we
demonstrate the ability to apply Kapitza-Dirac and Bragg
diffraction \cite{kozu99bragg} to cold molecules.  Using optical
standing waves of suitably chosen frequencies, sodium dimers were
coherently manipulated with negligible heating or other incoherent
processes.  First, we characterized the coherence of our ``source"
molecules, created via Feshbach resonance.  By measuring the Bragg
spectrum of the molecules immediately after their creation, the
conversion from atoms to molecules was shown to be coherent - the
matter wave analog to frequency doubling in optics. The quadratic
spatial dependence of the phase of the expanding molecules was
observed using an autocorrelation interference technique. By
creating a duplicate sample of molecules and overlapping it with
the original, matter wave interference was observed. Finally, the
matter wave analog to sum-frequency generation  was demonstrated.
Atoms prepared in two momentum states, prior to creating
molecules, were observed to cross-pair, generating a third
momentum state.

To produce molecules, sodium condensates were prepared in a
crossed optical dipole trap in the $|F,m_F\rangle=|1,-1\rangle$
state. Trap frequencies of ($\omega_x, \omega_y, \omega_z$) =
$2\pi \times $(146, 105, 23)~Hz yielded a typical peak density of
$n_0 \simeq 4.3 \times 10^{14}$~cm$^{-3}$ for 25 million atoms.
Atoms were then spin-flipped to the $|1,1\rangle$ state, in which
a 1~G wide Feshbach resonance exists at 907~G~\cite{inou98}.

The magnetic field sequence used to create and detect Na$_2$
molecules was identical to our previous work~\cite{xu03na_mol,
muka04}. Briefly, the axial (z-axis) magnetic field was ramped to
903~G in 100~ms. In order to prepare the condensate on the
negative scattering length side of the resonance, the field was
stepped up to 913~G as quickly as possible ($\sim$~1~$\mu$s) to
jump through the resonance with minimal atom loss. After waiting
1200 $\mu$s for the transient field to die down, the field was
ramped back down to 903~G in 50~$\mu$s to form molecules. In order
to remove non-paired atoms from the trap, the sample was
irradiated with a 10~$\mu$s pulse of resonant light. Because 903~G
is far from the Feshbach resonance, the mixing between atomic and
molecular states was small, and therefore molecules were
transparent to this ``blast'' pulse. By ramping the field back to
913~G, molecules were converted back to atoms. Absorption images
were taken at high fields (either at 903~G or 913~G), with the
imaging light incident along the axial direction of the
condensate.  Bragg scattering of atoms and molecules was carried
out using two nearly orthogonal beams ($\theta_B = 84^\circ$),
aligned such that particles were scattered along the x-axis of the
trap. The beams were far-detuned from atomic/molecular transitions
to limit spontaneous scattering.  For atoms the detuning was
$\simeq$ 4 nm from the sodium D lines.  To find a suitable
transition for the molecules, we scanned the laser wavelength and
measured the Rabi frequency for Bragg transitions. Several narrow
transitions were observed, but not carefully characterized. For
this work the laser was set to 590.159 nm and detuned $\sim$ 50
MHz from the apparent resonance, yielding a Rabi frequency of 2
kHz.

Fig \ref{fig:Figure1} shows time-of-flight images of Bragg
scattering for atoms and molecules.  Because the kinetic energy of
the scattered particles was much larger than their mean-field
energy, individual momentum states were well-resolved in ballistic
expansion. Both atoms and molecules receive equal two-photon
recoil momentum, $p_{r}= 2 h \sin(\theta_B/2)/\lambda_L$, where
$\lambda_L$ is is the wavelength of the Bragg beams. However,
scattered molecules move away from the central peak with half the
velocity of atoms, owing to their doubled mass. Fig
\ref{fig:Figure1}c,d show Kapitza-Dirac scattering, where multiple
atomic and molecular momentum states were populated due to the
broad frequency distribution of the short pulse (5 $\mu$s).

\begin{figure}[tbp]
\includegraphics[width=75mm]{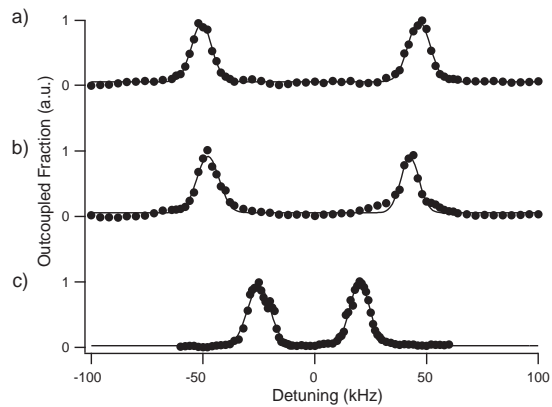} \caption{Bragg spectra
for atoms and molecules.  In (a) the spectrum is taken for a pure
atomic sample above the Feshbach resonance (913 G). (b) and (c)
are spectra of atoms and molecules, respectively, below the
resonance (903 G). In (c) atoms were removed from the trap with
resonant light to limit losses due to atom-molecule collisions.
The Bragg resonance condition for molecules occurs at half the
frequency of the atomic resonance.} \label{fig:Figure2}
\end{figure}

In order to study the coherence properties of the sample, Bragg
spectra \cite{sten99brag} were taken with $\sim$ 1 kHz resolution
by pulsing on the two laser beams (250 $\mu s$ square pulse)
before releasing the particles from the trap.  For noninteracting
particles, the Bragg resonance occurs at a relative detuning of
$\nu_0 = \pm p_{r}^2/2mh$ between the beams, where the sign of
$\nu_0$ dictates the direction of out-coupling. Interactions in a
condensate give rise to a mean-field shift of the resonance
$\delta\nu=4 n_0 U/7 h$, where U=$4 \pi \hbar^{2} a/m$ and $a$ is
the scattering length.  Fig \ref{fig:Figure2} shows three spectra
for (a) atoms above the Feshbach resonance, as well as (b) atoms
and (c) molecules after sweeping through the resonance. The
reduced mean-field shift for atoms below the resonance (Fig 1b)
can be attributed to inelastic losses caused by passing through
the resonance. Atoms below the resonance coexisted with a small
fraction of molecules (2$\%$). The peak output for each set of
data is normalized to 1. The actual peak out-coupled fractions
were 0.06 for the atoms and 0.5 for the molecules. The atomic
signal was kept intentionally low to minimize collisions, which
make the data analysis difficult (see halos in Fig 1c). As
expected from the resonance condition, molecular resonances occur
at half the frequency of atomic resonances.

Two mechanisms contribute to the fundamental width of the Bragg
resonance \cite{sten99brag,sten99bragerratum}.  For a parabolic
density distribution, the finite size of the sample yields a
momentum spread
\begin{equation}
    \Delta\nu_{p}  =  1.58\frac{p_{r}}{2\pi m x_{0}},
\end{equation}
where $x_0$ is the Thomas-Fermi radius. In addition, the
inhomogeneous density distribution of the sample produces a spread
in mean-field energy
\begin{equation}
    \Delta\nu_{n}  = \sqrt{\frac{8}{147}}\frac{n_0 U}{h}.
\end{equation}
The fundamental width is approximately given by the quadrature sum
of these two broadening mechanisms
$\Delta\nu=\sqrt{\Delta\nu_{n}^2+\Delta\nu_{p}^2}$.

The fundamental width, $\Delta\nu$, and measured rms width,
$\overline{\sigma}$, are compared for each case in Table I.  $n_0$
and $x_0$ were determined from the size of the condensate in
time-of-flight. The measured widths cannot be accounted for by
fundamental broadening mechanisms alone. For atoms above
resonance, the fundamental width is $\Delta\nu$=1.39 kHz, compared
to the measured value of 4.46 kHz. Therefore, another broadening
mechanism must contribute $\sim$ 4 kHz to the width. Most likely
this is due to Doppler broadening caused by random center-of-mass
motion and other collective excitations of the condensate. This
was investigated by mixing two frequencies into each Bragg beam to
out-couple particles in both $\pm$x directions. For particles at
rest, the out-coupling should always be symmetric. However, we
observe the ratio of out-coupled particles in either direction to
fluctuate. In addition we measure a small, consistent shift in the
Bragg spectrum, indicating a drift velocity. A line broadening of
4 kHz corresponds to a velocity of $\simeq$ 2 mm/s, or a
vibrational amplitude of $\simeq$ 2 $\mu$m (compared to $x_{0}= 13
\mu$m). This is not unreasonable, because the field ramping scheme
used to bring the atoms to high field is violent and may induce
collective excitations such as breathing modes.

\begin{table}[h]
\begin{tabular}
{|c |c |c|c|} \hline
Spectrum&$\Delta\nu$ (kHz)&$\overline{\sigma}$ (kHz)&T (nK)\\
\hline
Atoms (Above)&1.39\;\,&4.46(17)&10\\
Atoms (Below)&1.03\;\,&4.50(60)&10\\
Molecules &$0.36\,\ddag$&4.53(14)&20\\
\hline
\end{tabular}
\caption{Fundamental width ($\Delta\nu$), measured rms width
($\overline{\sigma}$), and the corresponding temperature (T),
assuming a thermal distribution.\\ \footnotesize{$^\ddag$This
lower bound assumes that the molecules have the same spatial
profile as the atoms, which our results indicate is not the
case.}}
\end{table}

Despite vibrational noise making the dominant contribution to the
width of the spectra, the measured values are still narrow enough
to indicate quantum degeneracy. For a given $\overline{\sigma}$,
the corresponding temperature for a thermal distribution of
particles is
\begin{figure}[h]
\includegraphics[width=35mm]{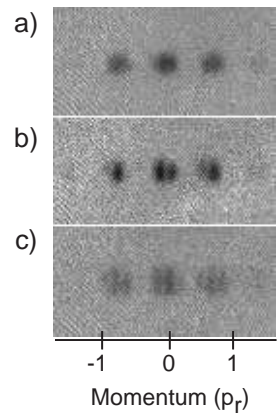}
\caption{Matter wave interference of molecules.  (a) The molecular
sample is split with a short Kapitza-Dirac pulse, creating two
identical copies with momenta $\pm p_{r}$. (b)-(c) After the
copies have separated for time $\Delta t$ = 100 $\mu$s, a second
pulse recombines them, giving rise to interference in each
momentum component. The time-of-flight in each image is 12 ms.}
\label{fig:Figure3}
\end{figure}
\begin{equation}
  T=\frac{m\, h^2\,\overline{\sigma}^2}{k_B \, p_r^2}
\end{equation}
Thus, for an rms width of 4.5 kHz, the temperature for a thermal
distribution of molecules would be 20 nK, comparable to a previous
value obtained using a time-of-flight technique \cite{xu03na_mol}.
The BEC transition temperature for our trap parameters
\cite{trap_parameters} and $5\times10^4$ molecules is much higher
(115 nK). This demonstrates a deeply degenerate, purely molecular
sample, where as previous experiments have demonstrated coherent
admixture of molecular character into an atomic BEC
\cite{donl02mol}.

The molecular Bragg Spectrum showed a surprisingly large shift of
$\delta\overline{\nu}=625$ Hz.  If interpreted as a mean-field
shift, this would imply either a very high density (possible due
to a spatial collapse) or an anomalously large molecular
scattering length outside the Feshbach resonance.  Further study
is necessary.

The spatial phase of the expanding molecular cloud was directly
imaged using an autocorrelation method \cite{sims00}, which gives
rise to the self-interference of the molecular sample (see Fig
\ref{fig:Figure3}).  To accomplish this, two identical copies of
the sample were made using a short Kapitza-Dirac pulse (10
$\mu$s), applied after 2 ms of ballistic expansion. The copies,
with momentum $\pm p_{r}$, moved away from the zero momentum peak
for time $\Delta t$ before an identical pulse recombined them with
the original. This type of interferometer has three readout ports,
with momenta 0, $\pm p_{r}$.  The straight-line interference
fringes are characteristic of a quadratic spatial phase. We
observe fringe contrast as high as 50\% and a fringe spacing
consistent with $\lambda_f=h t/m d$ \cite{andr97int}, where $d =
p_{r}\Delta t/m$ is the distance the copies moved between pulses
\cite{pulse}. Interference fringes can only be resolved for small
$d$. Therefore, this method cannot be used to observe coherence
lengths longer than those inferred from Bragg spectroscopy.  It
should be noted that the appearance of interference fringes does
not imply that the sample is condensed. Rather, it demonstrates
only that the coherence length in time-of-flight is longer than
the separation $d$. Therefore, similar interference can also be
observed for a cloud of thermal atoms \cite{bloc00coh,mill04}.

\begin{figure}[tbp]
\includegraphics[width=50mm]{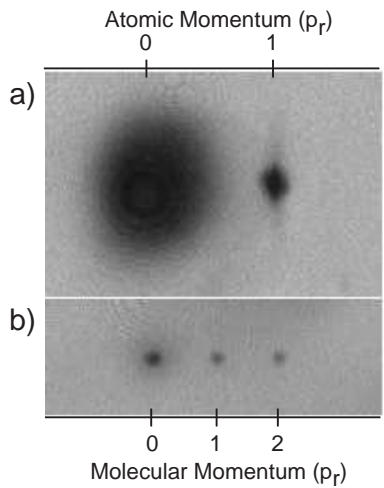} \caption{Sum frequency
generation of atomic matter waves.  (a) Atoms were initially
prepared in momentum states 0, 1. (b) By sweeping through the
Feshbach resonance, atoms combine to form molecules with momenta
0, 1, and 2.  Momentum state 1 is the sum frequency of the two
atomic matter waves. The ``nonlinear medium" is provided by the
atomic interactions.  The time-of-flight in each image is 17 ms.}
\label{fig:Figure4}
\end{figure}

The conversion of atoms to molecules may be viewed as the atom
optic equivalent of frequency doubling \cite{photonics}. The
relevant Hamiltonian for the atom-molecule coupling has the same
form as that for the optical frequency doubling process:
\begin{equation}
a_{2m}^{\dagger}a_m a_m
\end{equation}
where $a_m$ is the annihilation operator for the atomic field and
$a_{2m}^{\dagger}$ is the creation operator for the molecular
field. The measurement of the Bragg spectrum shows that the sharp
``linewidth" of the seed (atom) laser is inherited by the
molecular laser.  In nonlinear optics, photon interactions are
typically mediated by a refractive medium. Here, the nonlinearity
arises from the atoms themselves, in the form of s-wave
interactions.  The high density, or ``brightness", of the source,
together with the enhanced interactions at the Feshbach resonance
provide the means to combine two matter waves.  By combining two
disparate matter waves, rather than identical ones, we extend the
analogy of frequency doubling to the more general process of sum
frequency generation.  To do this, atoms were initially prepared
in momentum states 0, 1 (in units of $p_{r}$).  By sweeping
through the resonance, molecules were produced in three momentum
states: 0, 1, and 2 (see Fig. \ref{fig:Figure4}). States 0 and 2
are simply the frequency doubled components of the two initial
matter waves. State 1 however, results from cross pairing between
the initial momentum states, and is thus their sum frequency.
This is the first time that a Feshbach resonance was observed
between atoms colliding with a controlled non-vanishing momentum.
The Feshbach resonance should be slightly shifted compared to the
resonance for atoms at rest, which reflects the same physics
encountered in the temperature dependence of the position of the
resonance \cite{vladan99}.

In conclusion, we have demonstrated coherent molecular optics
using standing light waves.  The ability to coherently convert
atoms into molecules makes molecular optics even richer than atom
optics. In addition, the techniques demonstrated in this paper
could prove useful for probing molecules formed in
quantum-degenerate fermi systems, and possibly even Cooper pairs.

The authors would like to acknowledge M. Boyd and W. Setiawan for
experimental contributions and thank A.E. Leanhardt and M. W.
Zwierlein for critical readings of the manuscript. This research
is supported by NSF, ONR, ARO, and NASA.

\bibliographystyle{prsty}

\bibliography{MBReferences_condmat}

\begin{thebibliography}{10}
\bibitem{trap_parameters}
For an optical dipole trap using 1064 nm light, the trap parameters for both
  atoms and molecules are the same.
\bibitem{pulse}
Strictly speaking, this analysis only applies to the outcoupled ports. The
  center port is comprised of three copies.
\end{thebibliography}

\end{document}